\documentclass[prx,twocolumn,floatfix]{revtex4-1}
\usepackage{bbold}
\usepackage{graphicx}
\usepackage{amssymb}
\usepackage{amsmath}
\usepackage{empheq}
\usepackage{amsthm}
\usepackage{siunitx}

\usepackage{dcolumn}
\usepackage{bm}
\graphicspath{ {Figures/} }

\usepackage{color}
\usepackage{float}

\graphicspath{ {Figures/} }

\begin{document}


\title{Underwater Focussing of Sound by Umklapp Diffraction}

\author{Gregory~J.~Chaplain$^{1,2}$, Richard~V.~Craster$^{1,3,4}$, Nick Cole$^{2}$, Alastair P. Hibbins$^{2}$ and Timothy A. Starkey$^{2}$}

\affiliation{$^1$ Department of Mathematics, Imperial College London, London SW7 2AZ, UK\\
$^2$ Centre for Metamaterials Research and Innovation, University of Exeter, Exeter EX4 4QL, United Kingdom \\
$^3$ Department of Mechanical Engineering, Imperial College London, London SW7 2AZ, UK \\
$^4$ UMI 2004 Abraham de Moivre-CNRS, Imperial College London, London SW7 2AZ, UK}

\begin{abstract}
Scholte modes that are localized between a submerged axisymmetric-structured elastic plate and surrounding fluid can undergo mode conversion via Umklapp diffraction into radiative modes; this radiative response is  verified by experiments that show focussing of underwater sound across a broad range of frequencies. The diffracted beams, that form a cone, are engineered to exist at a desired spatial position, associated with an abrupt change in the patterning of the plate. These structures take the form of grooves present only on one side of the plate, yet the focussing phenomena is achieved on both sides, even as viewed from the flat surface. 
\end{abstract}

\maketitle

\section{Introduction}

Several classes of surface waves exist at the interfaces between differing media that propagate along the interface and exponentially decay away from it. In elasticity, surface Rayleigh waves \cite{rayleigh1885waves} and Lamb waves \cite{lamb1917waves} are common such examples. At interfaces between two elastic solids there exist localised Stoneley waves and radiative leaky Rayleigh waves \cite{stoneley1926effect,hess2002surface}; special cases of Stoneley waves exist at fluid-solid interfaces that are often called Scholte waves \cite{scholte1942stoneley,cegla2005material}. For thin elastic plates submerged in a fluid the Scholte modes on either interface can couple, displaying dispersive behaviour at low frequencies \cite{staples2021}. These interfacial waves are analogous to electromagnetic surface-plasmons \cite{reather1988surface} in that they exist independently of any periodic structuring. However, introducing a (typically sub-wavelength) periodic structure to the interfaces permits the existence of array-guided modes that exist, by virtue of the periodicity, through the interaction of evanescent diffracted fields from the structure (e.g perforations or cavities). These exist under many guises across several wave regimes; spoof surface plasmons in electromagnetism, between patterned metals and dielectrics \cite{pendry04a,hibbins2005experimental}; acoustic surface waves (ASW) in acoustics \cite{donato1978model,kelders1998ultrasonic}, often in air assuming sound hard rigid boundaries \cite{hou2007tuning}; 
Rayleigh-Bloch waves in elasticity, on structured half-spaces and thin plates \cite{porter99a,colquitt2015rayleigh,colquitt2017seismic,chaplain2019rayleigh}; and edge waves in water waves along coastlines \cite{evans93a}. Typical applications and phenomena for surface and array-guided waves range from sensing, energy harvesting \cite{chaplain2020delineating}, mode conversion \cite{skelton2018multi}, signal processing and material characterisation \cite{matikas2011new}. 

For the case of a patterned elastic plate submerged in a fluid there exist non-leaky acoustic surface waves that propagate in the fluid above the subwavelength structured surface \cite{he2011nonleaky,estrada2012engineering,hou2007tuning,graham2019underwater}, and it is the manipulation of these waves that we shall focus on throughout - specifically demonstrating the ability to couple these to radiative waves for focussing applications. Throughout this article we consider the `focussing' as a designed overlap of diffracted beams, similar to the focussing mechanism of Bessel beams \cite{mcgloin2005bessel}, rather than the focussing associated with an imaging system. In this sense, the focal spots presented throughout are regions in which acoustic energy is concentrated; this phenomenon only occurs due to the designs presented throughout.

\begin{figure}
    \centering
    \includegraphics[width = 0.5\textwidth]{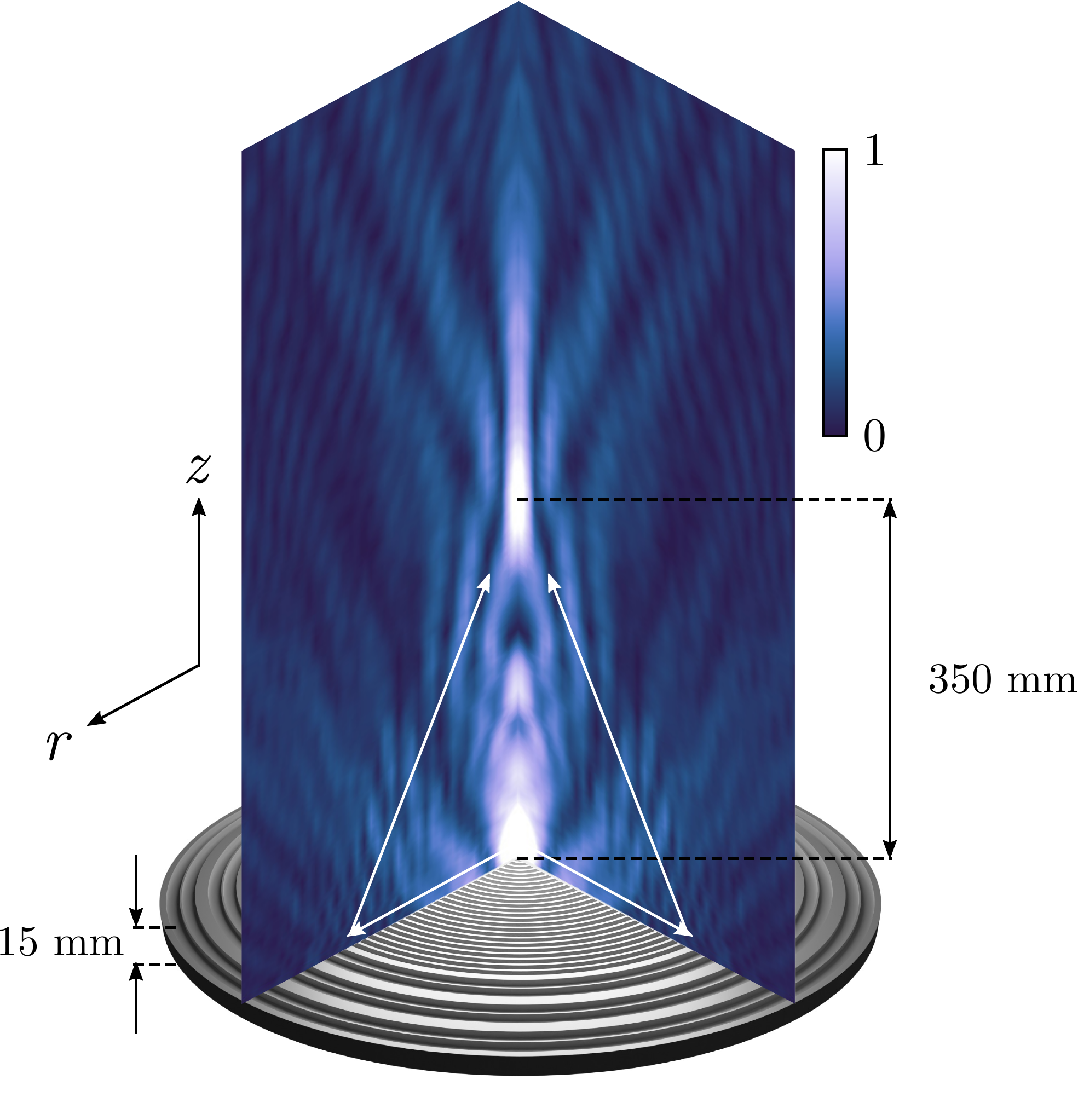}
    \caption{Focussing of underwater sound by Umklapp diffraction (experimental data, normalised absolute acoustic pressure): an acoustic source (56 kHz) located centrally beneath the plate (the lower surface is unstructured) is focussed to a spot within the fluid above the plate (structured side). White arrows show the ray path. A schematic of the plate is shown, with design given in Fig.~\ref{fig:R1}.}
    \label{fig:motivation}
\end{figure}

Here we demonstrate that localised Scholte modes can be modified (hereby termed modified Scholte modes (MSMs)) and mode converted to free-space acoustic radiation, and focussed over a broad frequency band. Our mechanism perturbs the dispersion of Scholte modes by adding concentric groove structures to an otherwise flat plate, modifying the supported symmetric and antisymmetric Scholte modes. We achieve focussing in free-space by designing a structure comprised of two differently patterned regions, each supporting their own MSM pair, with an abrupt transition between them. At this transition point, the diffraction of the MSMs results in the radiation of beams that can form a focal spot, the position of which is frequency dependent. We achieve this focussing over a broad frequency range ($\sim50~\si{\kilo\hertz}$).

Figure~\ref{fig:motivation} shows exemplar experimental data of this effect where an acoustic source ($56$ kHz) below the centre on the flat side of the plate is focused a distance above the plate on the structured side. The diffraction mechanism can be readily explained via considering diffractive scattering from higher Brillouin zones (BZ), i.e., Umklapp diffraction \cite{chaplain2020metasurfaces}. The theory presented for the modified Scholte modes is entirely general, and solely rests on the introduction of the designed periodic regions. As such it can be extended to manipulate, for example Lamb waves, and as such extend the frequency range of the effect.

The structure of this article is as follows. In Section~\ref{sec:theory} we outline the Umklapp diffraction mechanism with an illustrative example of a two-dimensional structured plate composed of two different one-dimensionally periodic regions and then, in Section~\ref{sec:design}, present the dispersion curves for both regions calculated using the Finite Element Method (FEM) with the commercial software COMSOL Multiphysics\textsuperscript{\textregistered}\cite{comsolacoustics}. To clearly and unequivocally visualise the focussing effect we use the one-dimensional periodic theory to inform the design of an axisymmetric device, this has the implication that the local curvature of the grooves that we now introduce is negligible and can be ignored. We justify this physically by only considering modes with circumferential order of 0, i.e we do not consider radially quantised modes of the device.
In Section~\ref{sec:ExpMethod} we outline the experimental method and detail the set-up. In Section~\ref{sec:ExpResults} we present the experimental results and cross-validation by comparisons with a completely bare, unstructured plate and finally draw conclusions in Section~\ref{sec:conc}.

\section{Theoretical Description}
\label{sec:theory}

Motivated by recent devices in electromagnetism and elasticity, which leverage so-called Umklapp diffraction \cite{chaplain2020ultrathin,chaplain2020tailored}, we apply these concepts to the setting of focussing underwater sound. 

The Umklapp or `flip-over' process, first hypothesised by Peierls \cite{Peierls_Thesis}, is conventionally used to understand the thermal conductivity at high temperatures due to phonon-phonon scattering \cite{Maznev2014}. The mechanism rests on the fact that wavevectors $\mathbf{k}$ in a periodic crystal are defined modulo a reciprocal lattice vector ${\mathbf{G}}$. The standard textbook definition of scattering events in a periodic crystal are then distinguished between normal N-processes and Umklapp U-processes according to
\begin{equation}
    \mathbf{k}_{1} + \mathbf{k}_{2} - \mathbf{k}_{3} = \begin{cases}
    \mathbf{0} & \text{N-process}, \\
    \mathbf{G} & \text{U-process}.
    \end{cases}
    \label{eq:umklapp}
\end{equation}

This definition then gives rise to the familiar reduced-zone scheme by translating wavevectors outwith the first BZ, effectively ‘folding’ the bands, by an integer multiple of reciprocal lattice vectors such that they lie within the first BZ. Despite its origins in solid state physics, not diffraction from surfaces, this (in the words of Peierls himself ``rather ugly" \cite{peierls2014bird}) term has been adopted to be synonymous with band folding effects \cite{foteinopoulou2005electromagnetic}. 

We utilise this phenomena to, at a designed spatial position, ensure that a propagating MSM with a wavevector in the first BZ of one periodic region excites a wave with wavevector outwith the first BZ in a second periodic structure. The resulting wave then undergoes Umklapp diffraction into the surrounding fluid with focussing achieved by the axisymmetry of the structure.

The general design paradigm to achieve this effect is as follows: (i) we first consider a thin, infinite elastic plate submerged in water. On one side of the plate we define a structuring that forms a 1D, infinitely periodic array of grooves, of pitch $a_1$ and depth $d$. The dispersion curves corresponding to the infinite crystal are then obtained, with this geometry being termed region 1 (R1). (ii) we then define, separately, a second infinite 1D array of grooves with a larger unit cell of pitch $a_2 > a_1$, such that the reciprocal space cell of this second region (henceforth R2) is smaller than that of R1. (iii)  The geometry of the second region is designed such that the supported MSM in R1 couples efficiently into R2; this requires matching the modal profiles in the regions and ensuring the dispersion curves of the two regions overlap. (iv) The two regions are then joined such that the boundary between them is abrupt (opposed to adiabatically graded e.g., \cite{chaplain2020delineating}). The dispersion curves of the infinitely periodic media are used to infer that a MSM in R1 will excite a MSM in R2; due to this wavevector now being beyond the first BZ in R2 it then couples to a radiative mode by diffraction through Umklapp scattering. Sound is diffracted at the transition point, at an angle determined by momentum conservation of the translated wavevector.

\begin{figure}
    \centering
    \includegraphics[width = 0.45\textwidth]{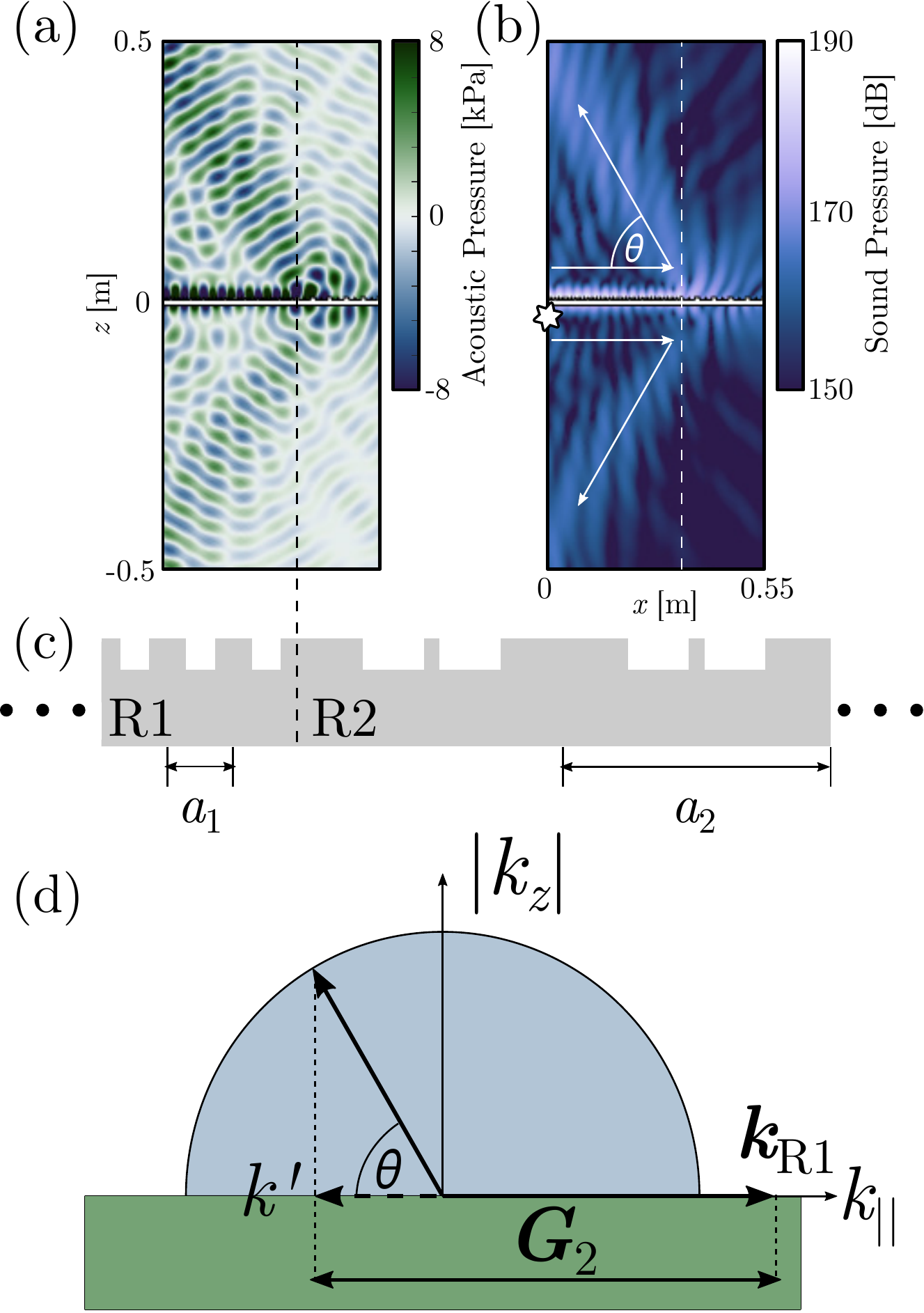}
    \caption{FEM simulation of Umklapp Diffraction at an interface between two regions of periodic structures. (a,b) Total acoustic pressure (amplitude and intensity  respectively), for a point source 1 mm below the flat side of the plate (29 kHz, marked as white star). The arrows in (b) show the ray path of the initial MSM in R1 which, upon reaching the boundary at R2 is diffracted out of the plate, forming beams with negative in-plane momentum. (c) A schematic illustrating the two regions (dashed vertical lines providing an indication the location of the change in structure). (d) Schematic of Umklapp diffraction mechanism whereby an incident surface wave, localised to the plate (green region), with wavevector $\boldsymbol{k}_{R1}$ is translated by a reciprocal lattice vector $\boldsymbol{G}_2$. The blue semicircle represents the isofrequency contour of the free fluid.}
    \label{fig:UmklappSlice}
\end{figure}

We elucidate this phenomena with an example, Fig.~\ref{fig:UmklappSlice}, which shows a plane of a FEM simulation with axial symmetry with absorbing boundary conditions on the edges of the domain. We structure one side of an elastic plate, partitioning it into two regions consisting of differing periodically-arranged grooves: R1 and R2, shown in Fig.~\ref{fig:UmklappSlice}(c). A MSM is excited in the first region using a point acoustic source (29 kHz) $1~\si{\milli\meter}$ below the centre of the plate on the lower, flat side. This MSM has in-plane wavevector $k_{||} = \boldsymbol{k}_{R1}$, shown in Fig.~\ref{fig:UmklappSlice}(d), which lies within the first Brillouin Zone of R1. At the interface between the two regions, due to the designed matching of mode shapes and overlap of the dispersion curves (Section~\ref{sec:design}), a MSM is excited in region R2. However, the corresponding wavevector now lies within the second BZ of R2 since we have ensured $a_{1} < a_{2} \implies \boldsymbol{k}_{R1} > X_{2}$, where $X_{2} \equiv k_{||} = \pi/a_2$. Due to the periodicity of R2 the wavevector of the excited MSM is equivalent to that translated by a reciprocal lattice vector, leaving its parallel component $k' = \boldsymbol{k}_{R1} - \boldsymbol{G}_{2}$ with $\boldsymbol{G}_2 = 2\pi/a_{2}$. This translated, or flipped, wavevector now lies within the isofrequency contour of the free fluid surrounding the plate, and hence couples to radiation within the bulk. The angle of the Umklapp diffracted beams is then determined from the relative orientation directions of the tangential component of the translated wavevector with that representing the sound cone.

In the following section we outline the exact design for the experimental verification of this effect, and the techniques used to obtain the dispersion curves for R2, which is crucial to the operation of the device.

\section{Design}
\label{sec:design}
\begin{figure}[b]
    \centering
    \includegraphics[width = 0.45\textwidth]{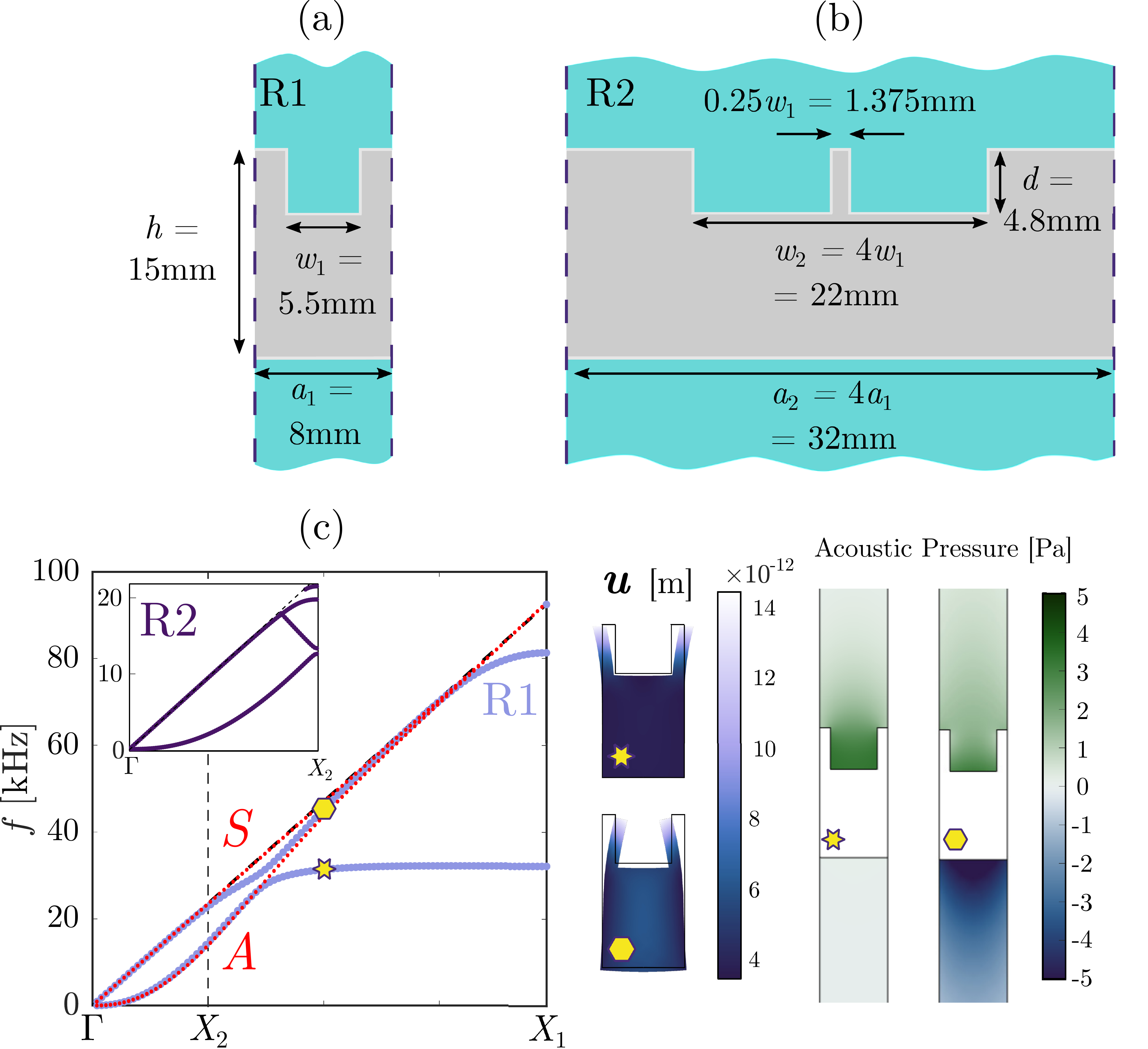}
    \caption{(a,b) Schematics of unit cells for R1 and R2 respectively; The aluminium plate is patterned with the grooves shown in each region, on one side only. Above and below the plate are regions of fluid (water). (c) Dispersion curves for the modes supported by a infinite array of the R1 unit cells (purple) and the coupled Scholte modes (antisymmetric ($A$) and Symmetric ($S$) respectively) for an unstructured plate (red). The inset similarly shows the modal dispersion curves in the first BZ for R2. The dashed black line shows the fluid sound line. The side panels show, at two marked frequencies, the plate displacement field $\boldsymbol{u} = (u,v)$ and the corresponding acoustic pressure fields in the fluid.}
    \label{fig:R1}
\end{figure}

We consider an aluminium plate, of thickness $h = 15~\si{\milli\meter}$, density $\rho = 2660~\si{\kilo\gram~\meter^{-3}}$ and Poisson's ratio $\nu = 0.34$ submerged in water, with density $\rho_f = 1000~\si{\kilo\gram~\meter^{-3}}$ and sound speed $c_f = 1480~\si{\meter\second^{-1}}$. The plate is split into the two periodic regions R1 and R2, the unit cells of which are shown in Figs.~\ref{fig:R1}(a,b) respectively. The unit cells comprising R1 are of pitch $a_1 = 8~\si{\milli\meter}$ with a groove of depth $d = 4.8~\si{\milli\meter}$ and width $w_1 = 5.5~\si{\milli\meter}$. The cells comprising R2 are such that $a_2 = 4a_1$ with  the groove width $w_2 = 4W_1$ equally partitioned by an additional pillar of width $w_1/4$, as shown in Fig.~\ref{fig:R1}(b). We then evaluate the band diagrams (dispersion curves) for an infinite, 1D periodic, array of unit cell that comprise R1 and R2. These give the frequencies of the supported mode as a function of wavevector parallel to the plate surface, $\boldsymbol{k}_{||}$, and are used to infer the behaviour of the axisymmetric device; consisting of the two regions rotated $2\pi$ about the device centre (Schematic shown in Fig.~\ref{fig:schem}). 

In order to calculate the dispersion curves of the localised MSM in each region, the FEM is used. To extract the decaying eigensolutions, a finite unit strip, of width $a_1$, is taken sufficiently long in the direction perpendicular to the periodicity such that the exponentially decaying solutions are not impacted by the boundary conditions at the top, or bottom, of the strip. Floquet--Bloch periodic conditions are applied to the side boundaries (dashed-lines in Fig.~\ref{fig:R1}). In doing so the dispersion curves within the first Brillouin-Zone, between $\Gamma-X_{1}$ ($X_{1}\equiv\boldsymbol{k_{||}} = \pi/a_1$), and $\Gamma-X_{2}$, $X_{2}\equiv \boldsymbol{k}_{||} = \pi/a_2$ are obtained, as shown in Fig.~\ref{fig:R1}(c). Shown too, at the frequencies marked by the star and hexagon, are example modal shapes of the structural displacement field  $\boldsymbol{u} = (u,v)$ through the exaggerated deformation of the solid structure. Also shown are the corresponding acoustic pressure fields in the fluid

Additionally shown in Fig.~\ref{fig:R1}(c) are the classical coupled Scholte mode dispersion curves (red points) obtained for a submerged unstructured plate with thickness $t = 10.2~\si{\milli\meter}$ i.e. a flat plate without the `pillars', opposed to the curves for a plate of thickness $t = 15~\si{\milli\meter}$ without the grooves of depth $d = 4.8~\si{\milli\meter}$). These are characterised by antisymmetric (A) and symmetric (S) branches, calculated numerically \cite{osborne1945transmission,kiefer2019calculating}. As expected, for a `hard' solid, where both sound speeds in the solid are greater than that in the liquid i.e. $c_{p} > c_{s} > c_{f}$, the symmetric coupled Scholte mode follows the sound line of the fluid. The effect of the structuring on these modes can be seen from the dispersion curves for the modes supported by R1; the lowest branch displays a flat band asymptote, characteristic of MSM resonance. This is evidenced by the confined acoustic pressure field on one side of the device (Fig.~\ref{fig:R1}). Unlike for simple sound-hard cavities the asymptotic frequency does not have a simple quarter-wavelength dependency \cite{graham2019underwater}. The upper branch of R1 above this resonance follows the dispersion relation of the asymmetric coupled Scholte mode. However, an important deviation from the classical Scholte dispersion can be seen at the band edge for the upper branch of R1, due to the periodic structuring. This affects the antisymmetric branch and can be seen through the antisymmetry of both the solid displacement and acoustic pressure fields on either interface. It is due to the modification of these modes by the influence of the periodic structuring we term these modified Scholte modes.

The dispersion curves of R2 within the first BZ are easily identified (inset of Fig.~\ref{fig:R1}(c)). However, to enable Umklapp diffraction we require dispersion curves in higher BZs (in this case up to the fourth BZ), which can be band-folded into the radiative regime within the first BZ. Extracting these eigen-solutions from the infinitely periodic unit-strip problem has the disadvantage that many spurious modes arise that are valid solutions only to the truncated numerical problem, but not the infinite physical one.

\begin{figure}[t]
    \centering
    \includegraphics[width = 0.45\textwidth]{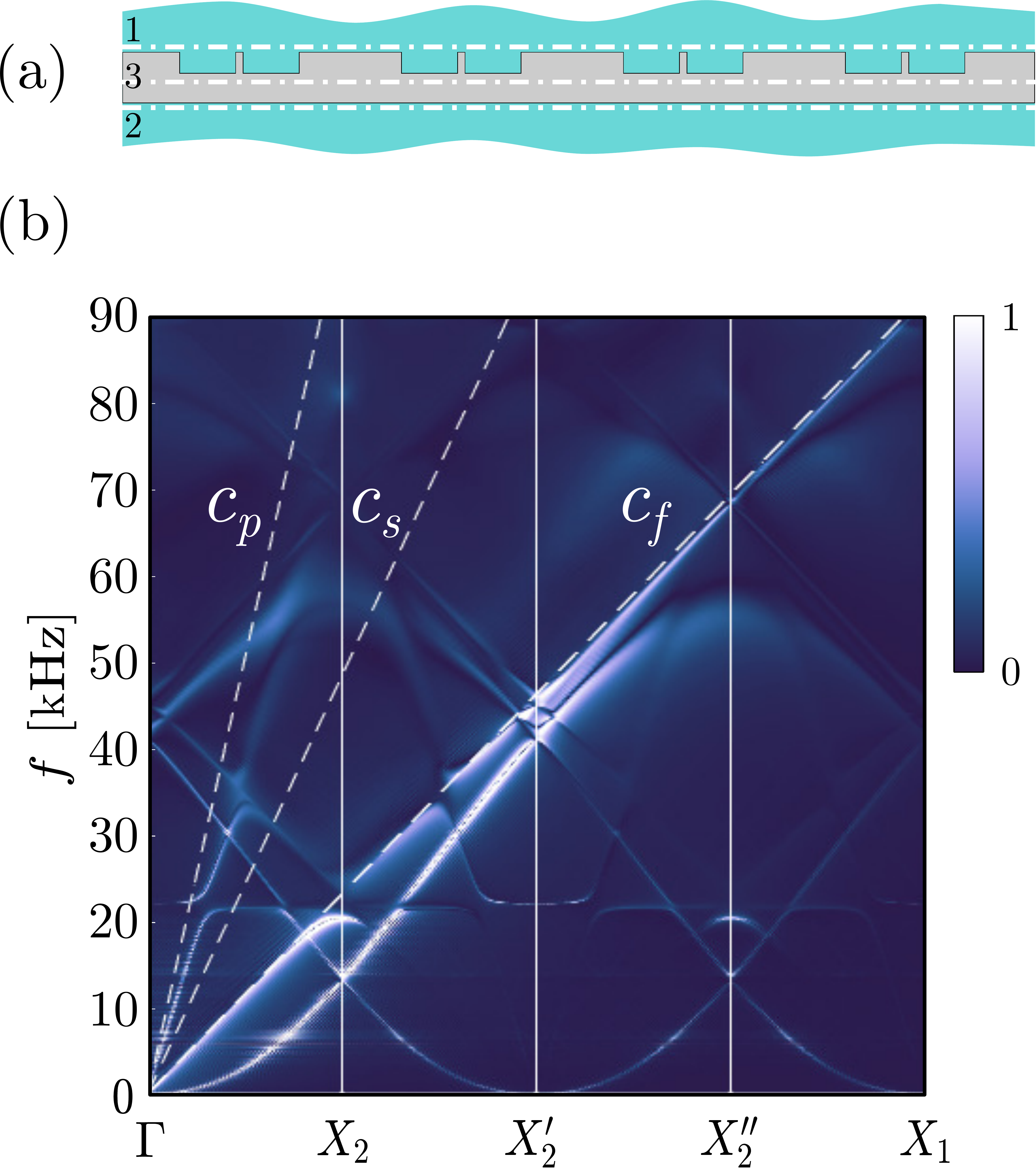}
    \caption{
Band diagram for R2. (a) Schematic of portion of numerical experiment domain. Dashed-dotted white lines labelled 1-3 show where the fields are extracted. (b) Normalised Fourier spectra of the acoustic pressure field along lines 1 and 2 and of the compressional and shear solid motion along line 3. The dashed white lines represent the three sound lines present: $c_f$ for the fluid sound line and $c_p$, $c_s$ for the elastic compressional and shear sound lines respectively.}
    \label{fig:R2}
\end{figure}

Several numerical schemes exist to extract the desired decaying solutions \cite{chaplain2019rayleigh,chaplain2020metasurfaces}, however the exercise remains tedious. Instead we opt to obtain the physical solutions via a numerical experiment by utilising the Fast Fourier Transform (FFT). This has the advantage of simplicity and, more importantly, provides additional information on the solid-fluid interaction. Instead of numerically solving the eigenvalue problem in the finite unit strip comprising R2, Fourier analysis is conducted on frequency domain simulations an array consisting only of R2. Fig.~\ref{fig:R2}(a) shows a schematic of a portion of such a domain. In total, 200 cells of R2 are taken, with absorbing boundary conditions placed on the extremities of the long array. MSMs are excited with a point monopole acoustic source placed $3~\si{\milli\meter}$ above the centre of the patterned surface. Performing a parametric sweep in frequency then gives the fields in both the fluid and solid domains. Along the dashed-dotted lines 1-2 shown in Fig.~\ref{fig:R2} the pressure field is extracted and the shear and compressional wavefields in the plate extracted along line 3. The spatial FFT is then computed, giving the required dispersion relations. 

In the band diagram for R2 (Fig.~\ref{fig:R2}) we see similar MSMs present in the higher BZs of R2; $X'_{2}$ marks the edge of the second BZ boundary (BZB) and $X''_{2}$ that of the third, with $X_{1} = 4X_{2}$ being the edge of the first BZ for R1, and the fourth BZB for R2. The presented dispersion curves, for each region, are for an infinite array of each unit cell (i.e. 1D-periodic). We now use these to infer the behaviour of an axisymmetric device consisting of 25 cells of R1 and three cells of R2. This cross section is rotated about the central axis of the plate, forming a disk consisting of concentric rings of R1 and R2, as shown in Fig.~\ref{fig:motivation}. We only consider isotropic MSMs with circumferential order 0 - we do not consider modes of the system with a higher degree of radial symmetry. As such the linear 1D dispersion curves provide a sufficient approximation to the supported modes of the axisymmetric device. 

Due to the careful design of the two periodic unit cells, over a large range of frequencies ($\sim 50~\si{\kilo\hertz}$), the dispersion curves of the two regions almost exactly overlap - see Fig.~\ref{fig:R2}. As such there is a very low impedance mismatch at the junction between the two regions; the excited parallel wavevector in R2 due to that in R1 are approximately equal i.e. $\Delta k_{||} = k^{R1}_{||}-k^{R2}_{||} \approx 0$. In addition to this the mode shapes possess the same symmetry in each region, and so the MSM in R2 is easily excited by the incoming MSM in R1. This means that at the transition region undesirable scattering is avoided, and so energy is primarily coupled into the designed negative diffractive orders through Umklapp scattering. This contributes to the high efficiency of the device.

We first detail the experimental procedure before presenting results and experimental corroboration. 

\section{Experimental Method}
\label{sec:ExpMethod}

\subsection{Sample fabrication and material properties}

The sample was fabricated based upon the unit-cell dimensions for regions 1 and 2 presented in Fig.~\ref{fig:R1}. From the plate centre, the sample comprises 25 cells of R1 and 3 cells of R2, which have axial symmetry about the centre of the plate rotated about the plate-normal as depicted schematically in Figs.~\ref{fig:motivation} and \ref{fig:UmklappSlice}(c).

The groove structures were milled using fluted-cutters with a CNC-assisted pillar drill into a circular aluminum-alloy (5083) plate mounted on a rotating table to accommodate this large radius sample. The plate sample has diameter, $D = 592 \pm 0.1$ mm and thickness, $h = 15.04 \pm 0.01 $ mm. Grooves have cavity depths, $d$, of  $5.07 \pm 0.05 $ mm (50 $\mu$m depth variation arises due to the flatness tolerance of the plate over the sample area). The thickness error in radial features of the surface relief profile, i.e. the pillar widths, were realised with much smaller positional tolerances of $\pm 0.005$ mm. For fabrication six 5 mm bolt holes and a 25 mm diameter, 6 mm deep recess in the plate centre were milled into the flat side of the sample (these are accounted for in the numerical simulations), the bolt holes were later used to mount the sample as shown in Fig. \ref{fig:schem} (b) and (c). A flat unstructured blank plate was also prepared as a reference.

The alloy plate has elastic material parameters: elastic modulus $E = 72.0 \pm 0.2~\si{\giga\pascal}$, shear modulus $G = 26.4 \pm 0.2 ~\si{\giga\pascal}$, density $\rho = 2660 \pm 610~\si{\kilo\gram~\meter^{-3}}$, and Poisson’s ratio $0.34 \pm 0.1$, as reported in the literature \cite{Holt} which are in agreement with recent acoustic characterisation measurements on submerged plates in this frequency regime \cite{graham2019underwater}.

\subsection{Acoustic measurements}

\begin{figure*}
    \centering
    \includegraphics[width = \textwidth]{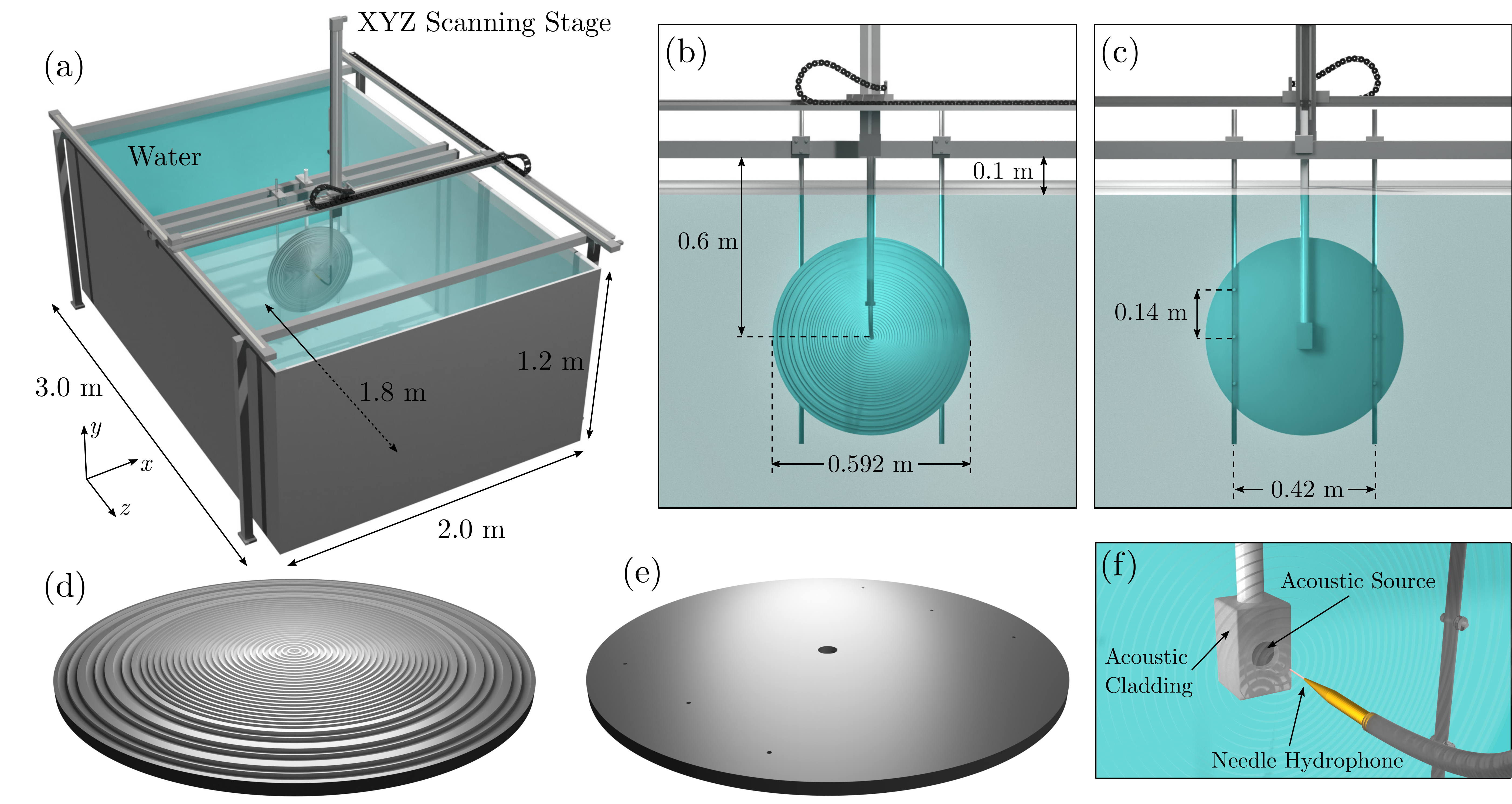}
    \caption{Schematic diagrams of tank arrangement and source/detector configurations. (a) render of water tank and $xyz$ scanning stage apparatus, with Umklapp sample suspended in place. (b) and (c) front and back views of the plate in the $xy$ plane for a flat-side excitation and structured-side detection configuration, and show the locations of the perspex mounts (d) and (e) rendered schematics of the sample's structured- and flat-side representative of the fabricated sample. (f) demonstration of the alignment between the shrouded transducer source and hydrophone detector at the plate centre, note in (f) the plate is made transparent to view the source.}
    \label{fig:schem}
\end{figure*}

To experimentally confirm acoustic focussing by our lens design, extensive time-gated acoustic characterisation of the fluid-field pressure distributions were made using a scanning tank facility (see Fig. \ref{fig:schem}(a)). Measurements were performed in a water tank without wall or surface treatments, with dimensions $3.0 \times 1.8 \times 1.2$ m ($L \times W \times D$). The sample was mounted to two perspex rods using bolts and washers to space the sample from the rod by 10 mm. It was then suspended on a cross bar so that the plate hung approximately at the half depth and width of the tank, and offset from the centre of the length to allow far-field measurements out to 1 m from the sample plate. Figure~\ref{fig:schem} (a)-(c) displays this experimental arrangement.

The sample was characterised under two different excitation conditions, when the source was positioned on the structured side, and when the source was on the flat side of the sample. For all experiments the sample was isonified using a ball-shaped Neptune-Sonar D70 transducer mounted at the centre of the sample approximately 5 mm from the surface (on either side of the sample). The transducer was shrouded in a foam enclosure to produce a 600 mm$^2$ area radiating aperture to produce a more point-like monopole acoustic excitation, and excited with a pulse with a 50 kHz centre frequency. This excitation condition provides a near-field acoustic source that excites over a range of incident angles, $\theta_i$, with the associated range of in-plane wavevectors, $k_\parallel$ ($= k_0 \sin {\theta_i} $), required for excitation of the modified Scholte modes. 

To obtain pressure field maps of sound radiated from the source and plate, the signal at the detection hydrophone (Brüel \& Kjær 8103 hydrophone) was scanned in space, using an $xyz$ scanning stage (in-house built with Aerotech controllers) to map the acoustic propagation; the voltage, $V$, from the detector was recorded as a function of time, $t$, at each position in the scan. 

At each spatial point, signals were averaged in time over 20 repeat pulses to improve the signal-to-noise ratio. The detector was sampled with sample rate, $f_s = 9.62$ MHz to record the signal for 1.6 ms at each point. The resulting usable frequency range for this source-detector response function was between 26 kHz and 90 kHz.

\section{Experimental Results}
\label{sec:ExpResults}

To illustrate the focussing arising from the tailored scattering of sound by the sample we present both temporal and frequency domain results.

\begin{figure*}
    \centering
    \includegraphics[width = \textwidth]{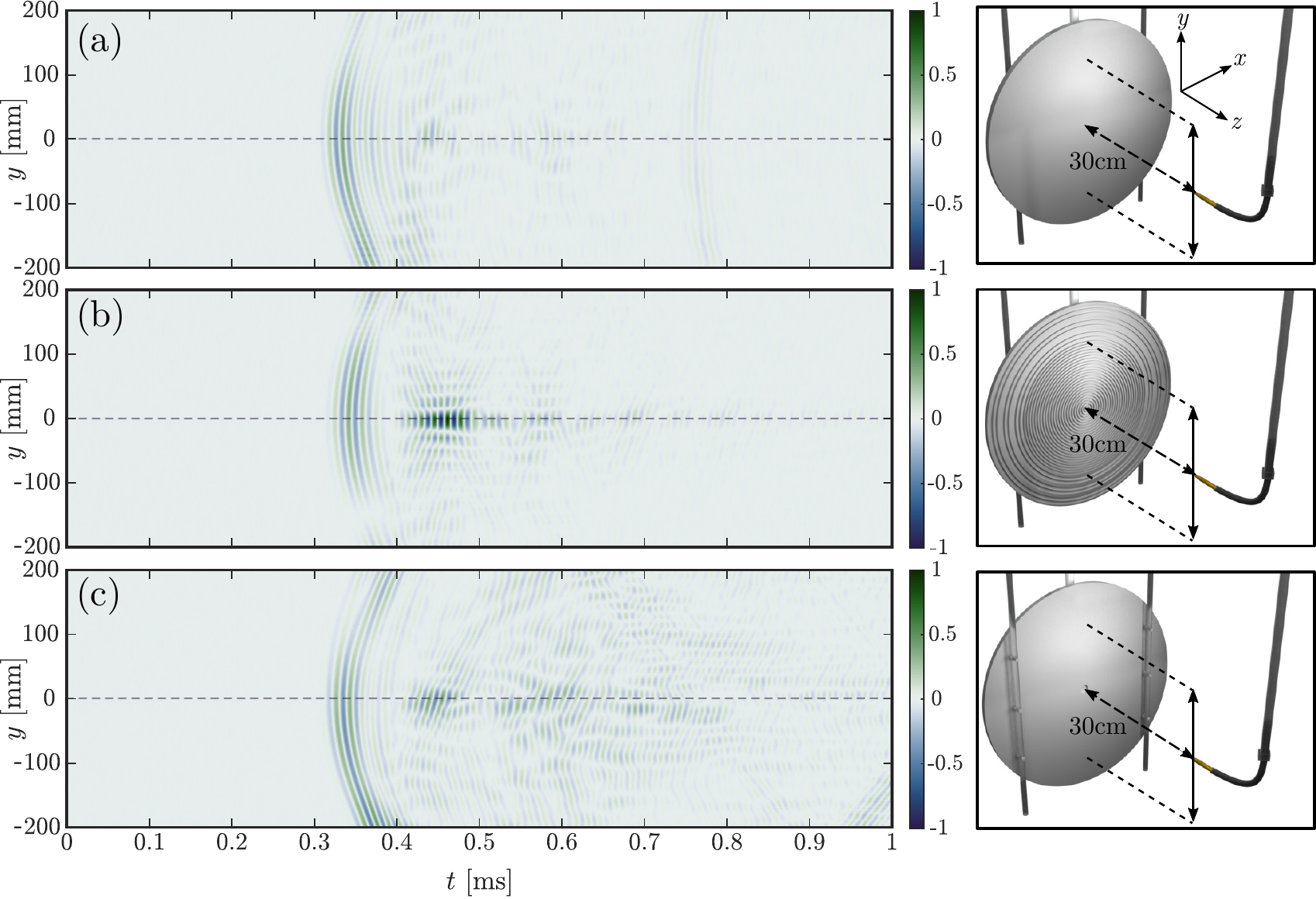}
    \caption{Time domain scans on a line parallel (at $ z = 300$ mm) to the sample in $y$-direction ($\pm 200$ mm relative to  $y = 0$ corresponding to the point normal to plate center).  Temporal maps show the acoustic signal for (a) an unstructured reference plate, (b) the Umklapp sample excited on the flat-side, detection on the structured side, and (c) the Umklapp sample excited on the structured side and measured on the flat side, as indicated with their associated schematics.}
    \label{fig:time_y}
\end{figure*}

As a consequence of our time-resolved excitation and detection we can first consider the arrival of the acoustic signal in the far field. Figure~\ref{fig:time_y} shows the arrival of sound as a function of time, spatially resolved along a line parallel to the sample at 300 mm in the $z$-direction. For these measurements a flat 15 mm thick reference plate (a) and the patterned sample (b) and (c), for two sample orientations, are measured for excitation on one side and detection on the opposite side of the sample.

The temporal evolutions display some common features; the leading edge of the pulse arrives at $\approx 0.32$ ms and disperses in $\pm y$ in a manner expected when intersecting a spherical wavefront, originating at the plate center, along a chord. In all cases the pulse displays the sinusoidal 3-cycle character of the input signal. For the unstructured reference plate (Fig. \ref{fig:time_y}(a)) the signal rings down and displayed the same spatial curvature. Other later arriving reflections are visible later in time, for instance at $\approx 0.76$ ms.

For the sample (excited on the flat side), in Fig. \ref{fig:time_y}(b), the behaviour is critically different; whilst the leading edge of the pulse behaviour is similar across the $y$-spatial scan, there is a significant signal enhancement later in time between $0.45 \lesssim t \lesssim 0.49$ ms at $y = 0$ indicating focussing along the plate normal from the centre. The implication is that the delayed signal has propagated along the plate before being scattered to free-space radiation at the designed transition between R1 and R2.

When the sample is flipped so excitation is now on the structured side, and detection on the flat side, as in Fig. \ref{fig:time_y}(c) a similar signal enhancement is seen. In this measurement, there is the presence of additional late arriving ($t \gtrsim 0.5$ ms) interference effects associated with unwanted diffraction by the perspex mounts that are now orientated on the detection side of the plate.

\begin{figure}
    \centering
    \includegraphics[width = 0.475\textwidth]{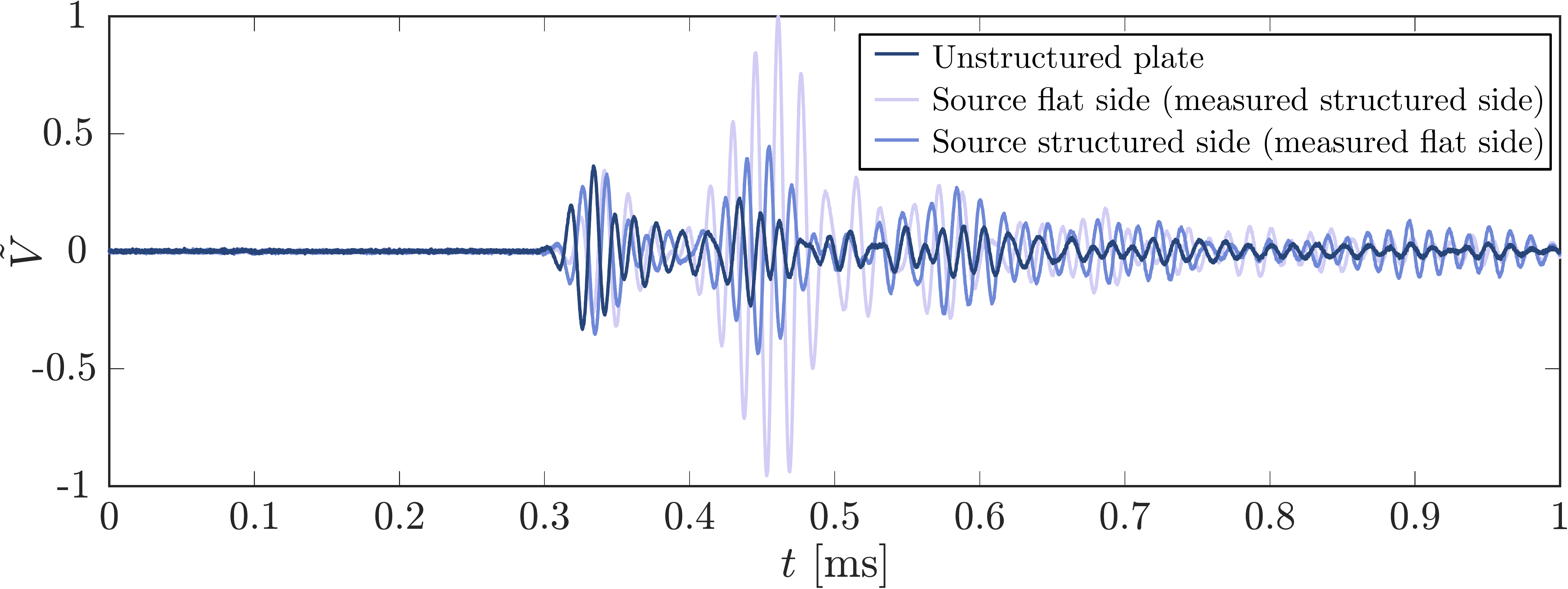}
    \caption{Time domain scans along centre lines ($y = 0$ mm) in Fig. \ref{fig:time_y}. The $y$-axis displays normalised voltage $\tilde{V}$ to maximum voltage in Fig. \ref{fig:time_y}(b). The relative measured wave-packet amplitude is over a factor of 2 larger in the patterned surface measurement then for the reference sample or the opposite structured-sample orientation.}
    \label{fig:time_y_centre}
\end{figure}

Closer inspection of the time domain signal shows that there is a significant enhancement of the voltage measured at the detector at $x, y, z = 0, 0, 300$ mm. The time signals for each configuration are plotted in Fig. \ref{fig:time_y_centre}, and show that (i) the relative measured wave-packet amplitude at the 'focus' is over a factor of 2 larger for the structured sample when excited on the flat face and measured on the patterned side when compared to the reference  sample or the opposite structured-sample orientation, and (ii) the amplitudes are essentially the same for the pulse that arrives first.

\begin{figure}
    \centering
    \includegraphics[width = 0.5\textwidth]{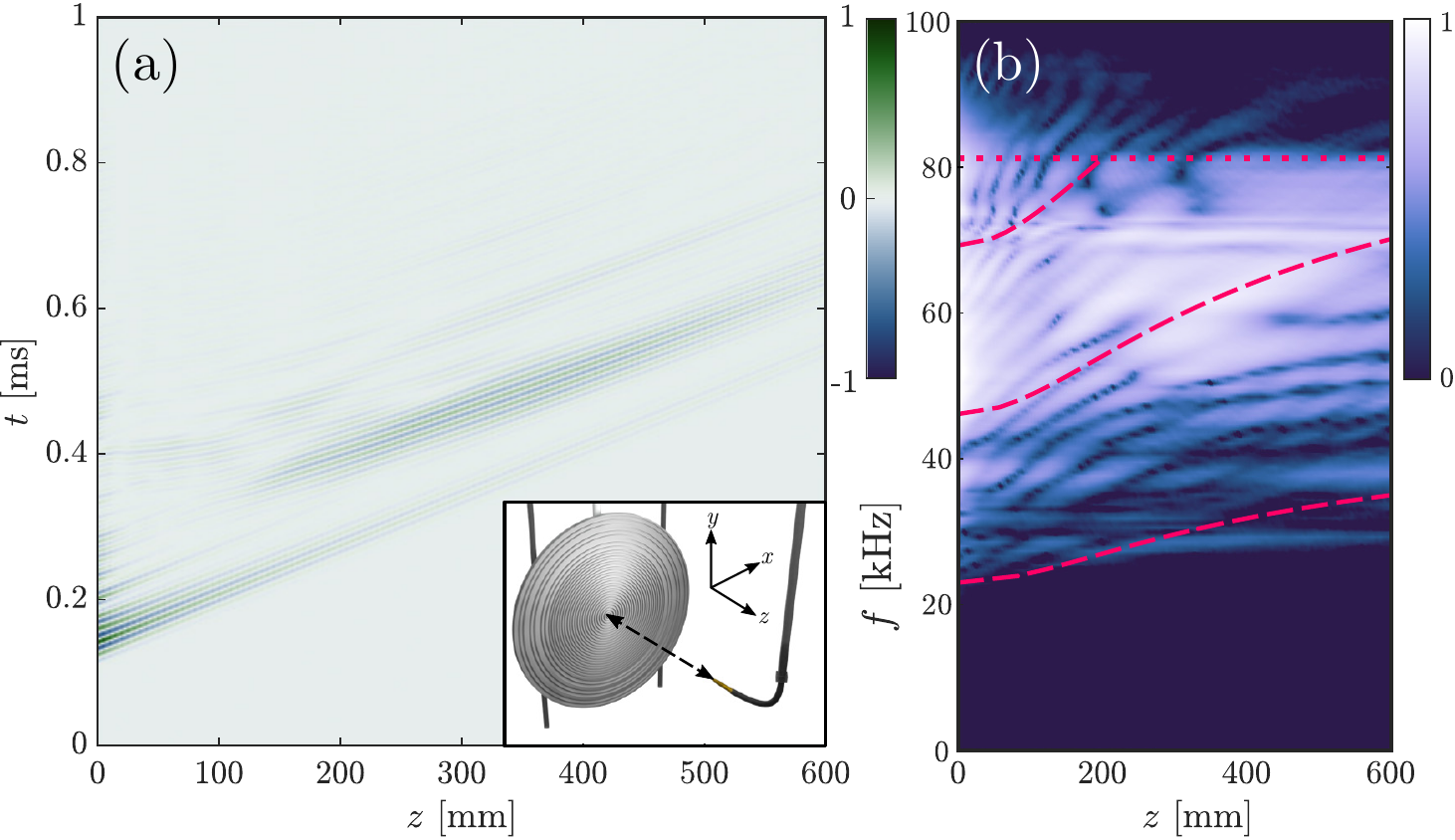}
    \caption{Linear detector scan measuring the temporal pressure variations at the plate centre in the plate normal (from $z$ = 0 to 600 mm). (a) Normalised time-domain results with inset indicating scanning direction. (b) Normalised logarithm of frequency-spectrum. Dashed lines show predicted focal spot position as function of frequency. Dotted line shows band-gap frequency for R1, above which no focal spots are observed as no guided wave is excited in R1.}
    \label{fig:time_z}
\end{figure}

Another way we explore this focal point is to observe the temporal evolution as a function of distance from the plate. Figure~\ref{fig:time_z}(a) shows a time domain scan along the $z$-direction normal to the plate centre. A focal spot is observed $\approx 500~\si{\milli\meter}$ from the plate at $\approx 0.5~\si{\milli\second}$. This time delay is consistent with the additional time of flight required for the guided wave to travel along the plate and travel the longer path length due to the angle of beaming. In Fig.~\ref{fig:time_z}(b) we show the frequency spectrum as a function of distance and make simple predictions of the focal spot position as a function of distance (dashed lines). These are evaluated simply by trigonometry using both the distance to the transition region ($25a_{1} = 200~\si{\milli\meter}$) and the angle of beaming. This is extracted from the dispersion curves by subtracting $m\boldsymbol{G}_{2}$, with $m\in 1,2,3$ (since we go up to the fourth BZ in R1) from the predicted $k_{||}$. The dotted line shows the band-gap frequency for R1, after which no focal spots are observed due to no MSM being excited in R1.

\begin{figure*}
    \centering
    \includegraphics[width = \textwidth]{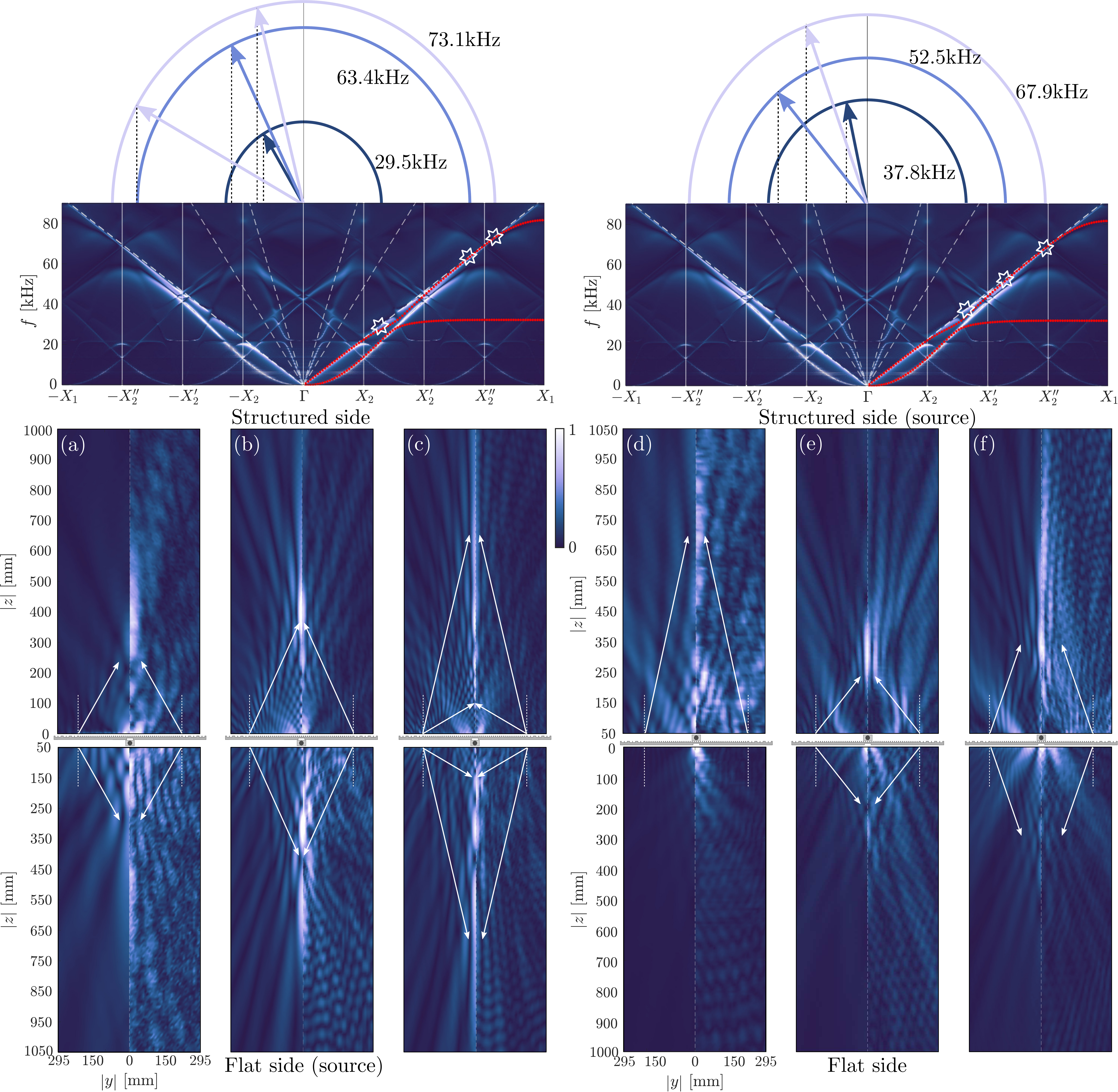}
    \caption{(Top left) \& (top right) numerical model band diagram analysis displaying the allowed scattering processes and resulting beaming angles for our experimental sample, for point-source acoustic excitation on the unstructured (top left) and structured (top right) when the pressure field is evaluated on the opposite side of the plate. Panels (a)-(f) display composite maps of the absolute pressure field ($|p|$) comparing numerical simulations and measurements. On each map the Umklapp sample and relative source location are indicated; the maps plotted above and below the plate are the simulated (on the left of the plate normal) and measured (on the right of the plate normal) absolute pressure fields. Maps (a)-(c) are for acoustic excitation on the flat-side of the sample for example frequencies (a) 29.5 kHz, (b) 63.4 kHz, and (c) 73.1 kHz. Maps (d)-(f) are for acoustic excitation on the structured-side of the sample for frequencies (d) 37.8 kHz, (e) 52.5 kHz, and (f) 67.9 kHz. Dash lines show the transition point from region 1 to region 2 on the sample, and the arrows show the expected angles at which sound will be projected, as calculated from the band-folding/scattering shown in the band diagrams (top left and right).}
    \label{fig:results}
\end{figure*}

Due to the physical constraints of the tank apparatus, all scans of the radiated fields were conducted in the $+z$ direction relative to the sample mount position, as indicated in Fig 5(a). The data presented required changing both the plate orientation and source position to produce the maps presented (i.e. two source positions x two sample orientations). In Fig.~\ref{fig:results} we show comparisons between simulations and experiments for a far field scan performed over a $yz$ half-plane, for both configurations of sources. We measure from $z = 0$ to $1~\si{\meter}$ when the source is on the opposite side of the plate than that being measured, and from $z = 0.005~\si{\meter}$ to $1.005~\si{meter}$ when we measure on the same side, to account for the proximity of the source and sample. In both cases we measure from $y = 0$ down to $y = - 0.3~\si{\meter}$ so to avoid encountering interference from diffraction off the tube holding the source (Fig.~\ref{fig:schem}(c)). Panels (a-c) and (d-f) in Fig.~\ref{fig:results} show comparisons of the simulated (left hand side) and experimental (right hand side) absolute pressure fields at several frequencies. The top left \& right panels show the dispersion curves for R1 (red) atop the full dispersion curves for R2 calculated via the numerical simulation. These are used, similarly to in Fig.~\ref{fig:UmklappSlice}, to predict the diffraction angle at the transition regions by subtracting multiples of $\boldsymbol{G}_{2}$ from the wavevectors at the frequencies marked by the white stars. In each case good agreement is seen between the theory, simulation and experimental results. 

The focussing of sound which we achieve results from the interference of diffracted beams from the transition region. The width of the beams is influenced by the number of cells in R2, and these beams spread as they travel through the free fluid. As such the size of the focal spot depends on the angle of the diffracted beams; the larger the angle the greater area of overlap of the spreading beams and hence the extension in $z$ of the focal `spot'. This is clearly visible when comparing, for example, Figs.~\ref{fig:results}(c) \& (e). 

Figures~\ref{fig:results} demonstrate it is therefore possible to couple non-radiative guided array modes, such as the modified Scholte modes here, to the surrounding bulk by engineering Umklapp diffraction at a given position. In each case, as expected, the diffraction is strongest on the structured side of the plate, and in general will depend on the momentum mismatch (i.e. the difference of $k_{||}$ in R1 and R2) and overlap in mode shape between each mode in R1 and R2. This is particularly noticeable in Fig.~\ref{fig:results}(d), which has the largest $\Delta k_{||}$.

\section{Conclusions}
\label{sec:conc}
We have demonstrated that the Umklapp diffraction method can be used to focus underwater sound over a broad range of frequencies. We achieved this by modifying the coupled-Scholte modes between a submerged elastic plate and the surrounding fluid by patterning only one side of the device with regions consisting of periodic concentric grooves. Comparisons to bare, unstructured plates confirm that the focussing capabilities are due to the designed structuring. The coupling between these supported guided waves to radiative modes is achieved by designing the regions to consist of unit cells which support similar mode shapes and have overlapping dispersion curves. Indeed the dispersion curves and band diagrams are key to the operation of the device and have been calculated through a combination of finite element methods and numerical experimentation, and highlight the need for careful design of the two comprising unit cells. We showed in simulation and experimentally verified that focal spots, whose position are frequency dependent, exist on both sides of the device, no matter which side the source is placed on. It is therefore possible to focus sound on the flat side of the device due to the coupling and modulation of the Scholte modes. 

The modulation of the non-radiative Scholte modes through periodic structuring is achieved through the general theory of Umklapp diffraction. As such we envisage extension to other wave types (e.g. Lamb waves) and anticipate applications for underwater sensing through non-destructive testing and evaluation, which is particularly relevant in the petrochemical industry. Further applications arise through the
acousto-fluidic control of micro-particles \cite{aubert2016simple}; the proposed structures could serve as the basis of hybrid bulk-acoustic-wave and surface-acoustic-wave devices \cite{friend2011microscale,cheeke2010fundamentals} and as such have potential reach in biochemical technologies.

\section*{Acknowledgements}
G.J.C gratefully acknowledges financial support from the EPSRC in the form of a Doctoral
Prize Fellowship, and from the Royal Commission for the Exhibition of 1851 in the form of a Research Fellowship. T.A.S. and A.P.H acknowledge financial support from DSTL. R.V.C thanks the EPSRC for support through grant EP/T002654/1.




%

\end{document}